\documentclass[12pt,a4paper]{article}%
\usepackage{epsfig}
\usepackage{amsmath}
\usepackage{amsfonts}
\usepackage{amssymb}
\usepackage{graphicx}%
\providecommand{\U}[1]{\protect\rule{.1in}{.1in}}
\begin{document}

\title{\textsc{Geometrical Origin of the Cosmological Constant}}
\author{\textsc{{\small H. Azri}}\thanks{E-mail: hm.azri@gmail.com} \text{ } and
\text{ } \textsc{{\small A. Bounames}}\thanks{E-mail: bounames@univ-jijel.dz}\\\textit{{\small Laboratory of Theoretical Physics, Department of Physics}},\\\textit{{\small University of Jijel, BP 98, Ouled Aissa, 18000 Jijel,
Algeria}}}
\date{}
\maketitle

\begin{abstract}
We show that the description of the space-time of general relativity as a
diagonal four dimensional submanifold immersed in an eight dimensional
hypercomplex manifold, in torsionless case, leads to a geometrical origin of
the cosmological constant. The cosmological constant appears naturally in the
new field equations and its expression is given as the norm of a four-vector
$U$, i.e., $\Lambda=6g_{\mu\nu}U^{\mu}U^{\nu}$ and where $U$ can be determined
from the Bianchi identities. Consequently, the cosmological constant is
space-time dependent, a Lorentz invariant scalar, and may be positive,
negative or null. The resulting energy momentum tensor of the dark energy
depends on the cosmological constant and its first derivative with respect to
the metric. As an application, we obtain the spherical solution for the field
equations. In cosmology, the modified Friedmann equations are proposed and a
condition on $\Lambda$ for an accelerating universe is deduced. For a
particular case of the vector $U$, we find a decaying cosmological constant
$\Lambda\propto a(t)^{-6\alpha}$. \newline


\end{abstract}

\newpage

\section{Introduction}

Since the discovery of the accelerating expansion of the universe
\cite{Riess1,Perlmutter,Knop,Riess2,Bennet}, dark energy is often invoked to
explain this phenomena and has led to renewed interest in the cosmological
constant. The simplest model for dark energy $(DE)$ is the cosmological
constant with the equation of state $\omega_{DE}=-1$. Nowadays, the
cosmological constant problem is one of the most fundamental problems in
physics \cite{Weinberg,Carrol,Padmanab,Peebles,Nobbenhuis}. Indeed, many
models with variable cosmological constant have been proposed, in some cases
it depends on space \cite{Narlikar,Tiwari,Dymnikova}, time \cite{Peebles88} or
both of them \cite{Sakharov}. Other authors have suggested that the
cosmological constant can be written as a trace of an energy-momentum tensor,
i.e., a Lorentz invariant scalar \cite{Rawaf,Raw
Taha,Abdel-rahm,Majern1,Majern2}.

In this paper, we try to tackle the cosmological constant problem from a
geometrical point of view. We will use Crumeyrolle's results on hypercomplex
manifolds where the space-time is considered as a diagonal four dimensional
submanifold immersed in an eight dimensional hypercomplex manifold
\cite{Crum1,Crum2,Crum3}. In his work, Crumeyrolle tried to obtain an unified
theory as that of Einstein-Schr\"odinger using a geometric construction in the
general case of a nonsymmetric connection, and the applications of this
approach have been performed by Clerc \cite{Clerc1,Clerc2,Clerc3}. A similar
approach has been suggested where the tangent bundle of space-time is endowed
with a hypercomplex algebraic structure \cite{Kunstatter83,Moffat84,Mann85}.
Historically, Einstein was the first who used a complex metric in order to
unify gravity and electromagnetism \cite{Einstein1,Einstein2}. Moffat has also
used a nonsymmetric complex metric as an attempt to a new theory of gravity
\cite{Moffat79}. In the last decades, we note that complex and hypercomplex
coordinates have been used in both field theory \cite{Schuller,Greiner1}, and
general relativity \cite{Lovelock1,Lovelock2,Chams,Mantz1,Greiner2,Greiner3}.
In the literature, the hypercomplex numbers are also called complex hyperbolic
numbers, pseudo-complex numbers, double numbers, paracomplex numbers or split
numbers
\cite{Schuller,Hucks,Yaglom,Para,Kantor,Ulrych1,Ulrych2,Fjelstad,Capelli,Olariu,Antonuccio}

On the other hand, it is important to note that there are also other geometric
approaches especially those accounting for the acceleration of the universe by
modifying the Einstein-Hilbert action \cite{vol}, and a truly geometric
approach to modified gravity without affecting the early universe
\cite{pun,schu1}. We note that there is a clear physical principle, discovered
from an investigation of the dynamical symmetries of Born-Infeld
electrodynamics, motivating the consideration of an eight-dimensional
pseudo-complex spacetime and which is related to the maximal acceleration
\cite{schu2,schu3}.

In this work, we apply the results of Crumeyrolle to the torsionless case in
order to describe the theory of general relativity. The Ricci scalar in the
space-time submanifold is given by $P=R+\Lambda$ where $R$ is the Ricci scalar
of the Levi-Civita connection and $\Lambda$ is a scalar function of a
four-vector $U$, and therefore has a geometrical origin due to the immersion
of the space-time submanifold in an eight dimensional manifold. It appears as
a correction of the Ricci scalar curvature $R$ used in the general relativity
\cite{Clerc1,Clerc2}. Using a variational principle, the modified Einstein's
equations are obtained where $\Lambda$ is identified to be the cosmological
constant. From the Bianchi identities, a particular form of the four vector
$U$ can be given, this leads to a decaying cosmological constant
$\Lambda\propto a(t)^{-6\alpha}$ which gives a correction to the time
evolutions of the matter and radiation as can be seen from the conservation law.

The main result of this paper is that the cosmological constant appears
naturally in the modified Einstein's equations as a norm of a four-vector $U$,
$\Lambda=6g_{\mu\nu}U^{\mu}U^{\nu}$. Consequently the cosmological constant is
space-time dependent, Lorentz invariant, and may be positive, negative or
null, depending on the nature of the four-vector $U$. The resulting energy
momentum tensor of the dark energy depends on the cosmological constant and
its first derivative with respect to the metric $g^{\mu\nu}$. From the
modified Friedmann equations, we deduce a condition on the cosmological
constant for an accelerating universe.

This paper is organized as follows: In section II, we recall briefly
Crumeyrolle's work on which our paper is based. In section III, we derive the
modified Einstein's equations using a variational principle. Then, we derive
the spherical solution in section IV and discuss its limit. In section V, the
modified Friedmann equations are deduced and the results are discussed. We
apply our result in section VI for a special form of the four vector $U$ and
deduce a decaying cosmological constant. Finally, the conclusion is given in
section VII.

\section{ Space-time as a diagonal submanifold}

In Crumeyrolle's work \cite{Crum1,Crum2,Crum3}, the space-time $V_{4}^{\text{
\ }}$ is considered as a diagonal submanifold of a $C^{\infty\text{ \ }}$
eight dimensional manifold $V_{8}^{\text{{}}}$ which is the product of two
identical four dimensional real manifolds $W_{4}^{\text{ }}$
\begin{equation}
V_{8}^{\text{ }}=W_{4}^{\text{ }}\times W_{4}^{\text{ }}. \label{stru}%
\end{equation}
The notation adopted in this work is : in $V_{8}^{\text{{}}}$ we use Latin
indices $(i,j,..=1,...8)$ and the operation $\ast$ on Latin indices is defined
such that $i^{\ast}=i\pm4,$ and in $V_{4}^{\text{ \ }}$ we use Greek indices
$(\alpha,\beta,..=1,...4)$ and $(\alpha^{\ast},\beta^{\ast}=1^{\ast
},...,4^{\ast})$. The construction (\ref{stru}) confers to $V_{8}$ a
hypercomplex structure. Indeed, if we use the hypercomplex coordinates
$X^{\alpha}=x^{\alpha}+Ix^{\alpha^{\ast}}$ where $I^{2}=1$, $\left(
x^{\alpha},x^{\alpha^{\ast}}\right)  $ are real coordinates from $W_{4}\times
W_{4}$, the diagonal submanifold $V_{4}^{\text{ \ }}$ is equivalent to
\cite{Crum1,Crum2,Crum3}
\begin{equation}
x^{\alpha^{\ast}}=0.
\end{equation}
The real coordinates $\left(  x^{\alpha},x^{\alpha^{\ast}}\right)  $ are
called the associated diagonal coordinates.

Suppose that the manifold $V_{8}^{\text{ }}$ is endowed with a symmetric non
degenerate metric tensor $\mathcal{G}_{_{ij}}$, then $V_{8}^{\text{{}}}$ is
seen to have a structure of a pseudo-riemannian manifold. According to
$\mathcal{G}_{ij}$, the metric tensor $g_{_{\alpha\beta}}$, in $V_{4}%
^{\text{{}}}$ is defined by setting in the natural diagonal frames of
$V_{4}^{\text{ }}$ (intrinsic conditions) \cite{Crum1,Crum2,Crum3}%
\begin{equation}
\widehat{\mathcal{G}}_{_{\alpha\beta}}=\widehat{\mathcal{G}}_{_{\alpha^{\ast
}\beta^{\ast}}}=0,\text{ }\ \ \ \ \ \widehat{\mathcal{G}}_{_{\alpha\beta
^{\ast}}}=\widehat{\mathcal{G}}_{_{\beta^{\ast}\alpha}}=g_{_{\alpha\beta}%
}\text{ }, \label{d10}%
\end{equation}
where $\widehat{}$ means the restriction in $V_{4}^{\text{ }}.$

A generalization of the Ricci theorem is to postulate that for every path of
$V_{4}^{\text{ }}$, the covariant derivative of the tensor $\mathcal{G}%
_{_{ij}}$ vanishes%

\begin{equation}
\nabla_{\rho}\mathcal{G}_{_{ij}}=0,\text{\ \ \ \ \ }\label{d11}%
\end{equation}
using conditions $($\ref{d10}$)$, the above relations (\ref{d11}) will be
written in $V_{4}$ as
\begin{equation}
\nabla_{\rho}\widehat{\mathcal{G}}_{_{\alpha\beta}}=\nabla_{\rho}%
\widehat{\mathcal{G}}_{_{\alpha\beta^{\ast}}}=\nabla_{\rho}\widehat
{\mathcal{G}}_{_{\alpha^{\ast}\beta^{\ast}}}=0.\label{d12}%
\end{equation}
Consider the connections in the natural diagonal frame bundle of $V_{8}$ such
that \cite{Crum1,Crum2,Crum3,Clerc2}%
\begin{equation}
\Gamma_{jk}^{i}=\Gamma_{j^{\ast}k}^{i^{\ast}}\text{ },\text{ \ \ }\Gamma
_{jk}^{i}=\Gamma_{j^{\ast}k^{\ast}}^{i}\text{ }.
\end{equation}
By restriction in $V_{4}^{\text{ }}$, we obtain
\begin{equation}
\Gamma_{\beta\gamma}^{\alpha}=\Gamma_{\beta^{\ast}\gamma}^{\alpha^{\ast}%
}=\Gamma_{\beta\gamma^{\ast}}^{\alpha^{\ast}}=\Gamma_{\beta^{\ast}\gamma
^{\ast}}^{\alpha},\text{ \ \ \ \ }\Gamma_{\beta\gamma}^{\alpha^{\ast}}%
=\Gamma_{\beta^{\ast}\gamma}^{\alpha}=\Gamma_{\beta\gamma^{\ast}}^{\alpha
}=\Gamma_{\beta^{\ast}\gamma^{\ast}}^{\alpha^{\ast}}\text{ },\label{D7}%
\end{equation}
we can show that the coefficients $\Gamma_{jk}^{i}$ with even number of
asterisks transform as connections, while the others (with odd number of
asterisks) transform as tensors in all natural diagonal frame of
$V_{4}^{\text{ }}$ \cite{Yano}.

Then according to Eqs.$(\ref{D7})$, one can define in $V_{4}^{\text{ }}$ a
connection $\mathcal{L}_{\gamma\beta}^{\alpha}$ and a tensor $\Lambda
_{\beta\gamma}^{\alpha\text{ }}$ by the relations \cite{Clerc2}%

\begin{equation}
\Gamma_{\beta\gamma}^{\alpha}=\Gamma_{\beta^{\ast}\gamma}^{\alpha^{\ast}%
}=\Gamma_{\beta\gamma^{\ast}}^{\alpha^{\ast}}=\Gamma_{\beta^{\ast}\gamma
^{\ast}}^{\alpha}=\mathcal{L}_{\gamma\beta}^{\alpha},\text{ \ \ \ \ }%
\Gamma_{\beta\gamma}^{\alpha^{\ast}}=\Gamma_{\beta^{\ast}\gamma}^{\alpha
}=\Gamma_{\beta\gamma^{\ast}}^{\alpha}=\Gamma_{\beta^{\ast}\gamma^{\ast}%
}^{\alpha^{\ast}}=\Lambda_{\beta\gamma}^{\alpha}. \label{D8}%
\end{equation}
The connection $\mathcal{L}_{\beta\gamma}^{\alpha}$ is generally nonsymmetric.

Using the relations $\left(  \ref{d10}\right)  $ and the properties
$(\ref{D8})$, the conditions $\left(  \ref{d12}\right)  $ give
\begin{align}
\partial_{\rho}g_{\alpha\beta}-\mathcal{L}_{\rho\alpha}^{\gamma}g_{\gamma
\beta}-\mathcal{L}_{\rho\beta}^{\gamma}g_{\alpha\gamma}  &  =0,\label{D13}\\
\Lambda_{\alpha\lambda}^{\gamma}\text{ }g_{\gamma\beta}+\Lambda_{\beta\lambda
}^{\gamma}\text{ }g_{\gamma\alpha}  &  =0,\text{ \ \ }\label{D12}\\
\Lambda_{\alpha\lambda}^{\gamma}\text{ }g_{\beta\gamma}+\Lambda_{\beta\lambda
}^{\gamma}\text{ }g_{\alpha\gamma}  &  =0. \label{D122}%
\end{align}
The solution of equation $(\ref{D13})$ is the general nonsymmetric connection
\cite{Clerc2}
\begin{equation}
\mathcal{L}_{\beta\gamma}^{\alpha}=\left\{  _{\beta\gamma}^{\alpha}\right\}
+g^{\alpha\rho}\left(  g_{_{\beta\tau}}S_{_{\rho\gamma}}^{\tau}+g_{_{\gamma
\tau}}S_{_{\rho\beta}}^{\tau}\right)  +S_{_{\beta\gamma}}^{\alpha},
\label{D14}%
\end{equation}
where $\left\{  _{\beta\gamma}^{\alpha}\right\}  =\frac{1}{2}g^{\alpha\rho
}\left(  \partial_{\beta}g_{\rho\gamma}+\partial_{\gamma}g_{\beta\rho
}-\partial_{\rho}g_{\gamma\beta}\right)  $ is the usual Christoffel symbol and
$S_{\gamma\beta}^{\sigma}$ is the torsion tensor.

The solution of equations $(\ref{D12})$ and $(\ref{D122})$ is the
antisymmetric tensor $\Lambda_{_{\beta\alpha}}^{\sigma}$ in $\alpha,\beta$%

\begin{equation}
\Lambda_{_{\beta\alpha}}^{\sigma}=g^{\sigma\gamma}\epsilon_{\gamma\beta
\alpha\rho}U^{\rho}, \label{D15}%
\end{equation}
where $\epsilon_{\gamma\beta\alpha\rho}$ is the antisymmetric Levi-Civita
tensor, and $U^{\rho}$ is an arbitrary $4$-vector in $V_{4}$.

By the immersion of the submanifold $V_{4}^{\text{ }}$ in the manifold
$V_{8\text{ }}^{\text{ }}$, the curvature form induced in $V_{4}^{\text{ }}$
is \cite{Clerc2}%
\begin{equation}
\widehat{\Omega}_{j}^{i}=\frac{1}{2}\widehat{R}_{j\lambda\mu}^{i}\text{
}dx^{\lambda}\wedge dx^{\mu},
\end{equation}
where $\widehat{}$ \ means the restriction in $V_{4}^{\text{ }}$ (remember
that $x^{\mu^{\ast}}=0$ \ in $V_{4}^{\text{ }}$)$.$

Then the induced curvature tensor in $V_{4}^{\text{ }}$ becomes
\begin{equation}
\widehat{R_{j\lambda\mu}^{i}}=\partial_{\lambda}\text{ }\Gamma_{j\mu}%
^{i}-\partial_{\mu}\text{ }\Gamma_{j\lambda}^{i}+\Gamma_{\rho\lambda}^{i}
\Gamma_{j\mu}^{\rho}\text{}+\Gamma_{j\mu}^{\rho^{\ast}}\text{ }\Gamma
_{\rho^{\ast}\lambda}^{i}-\Gamma_{\rho\mu}^{i}\Gamma_{j\lambda}^{\rho}\text{ }
-\Gamma_{\rho^{\ast}\mu}^{i}\Gamma_{j\lambda}^{\rho^{\ast}}\text{ },
\label{D19}%
\end{equation}

by contraction and using Eqs.$\left(  \ref{D8}\right)  $, one can obtain two
independent Ricci tensors in $V_{4}^{\text{ }}$ \cite{Clerc2}
\begin{align}
P_{\alpha\beta}  &  =\widehat{R_{\beta\lambda\alpha}^{\lambda}}=\partial
_{\lambda}\mathcal{L}_{\alpha\beta}^{\lambda}-\partial_{\alpha}\mathcal{L}%
_{\lambda\beta}^{\lambda}+\mathcal{L}_{\lambda\rho}^{\lambda}\mathcal{L}%
_{\alpha\beta}^{\rho}-\mathcal{L}_{\alpha\rho}^{\lambda}\mathcal{L}%
_{\lambda\beta}^{\rho}+\Lambda_{\rho\lambda}^{\lambda}\Lambda_{\beta\alpha
}^{\rho}-\Lambda_{\rho\alpha}^{\lambda\text{ }}\Lambda_{\beta\lambda}^{\rho
},\label{V1}\\
Q_{\alpha\beta}  &  =\text{ }\widehat{R_{\alpha^{\ast}\lambda\beta}^{\lambda}%
}=\partial_{\lambda}\Lambda_{\alpha\beta}^{\lambda}-\partial_{\beta}%
\Lambda_{\alpha\lambda}^{\lambda}+\mathcal{L}_{\lambda\rho}^{\lambda}%
\Lambda_{\alpha\beta}^{\rho}-\mathcal{L}_{\beta\rho}^{\lambda}\Lambda
_{\alpha\lambda}^{\rho}+\Lambda_{\rho\lambda}^{\lambda}\mathcal{L}%
_{\beta\alpha}^{\rho}-\Lambda_{\rho\beta}^{\lambda}\mathcal{L}_{\lambda\alpha
}^{\rho}\text{ }.
\end{align}

The two other Ricci tensors $\overline{P}_{\beta\alpha}$ and $\overline
{Q}_{\beta\alpha}$ can be obtained from $P_{\alpha\beta}$ and $Q_{\alpha\beta
}$ using Einstein's principle of pseudo-hermiticity \cite{Clerc2}, i.e., by
changing $\mathcal{L}_{\beta\gamma}^{\alpha},$ $\Lambda_{\beta\gamma}^{\alpha
},$ $g_{\alpha\beta}$ by $\mathcal{L}_{\gamma\beta}^{\alpha},$ $\Lambda
_{\gamma\beta}^{\alpha}$, $g_{\beta\alpha}$ in $P_{\alpha\beta}$ and
$Q_{\alpha\beta}$ respectively
\begin{align}
\overline{P}_{\beta\alpha}  &  =\partial_{\lambda}\mathcal{L}_{\alpha\beta
}^{\lambda}-\partial_{\beta}\mathcal{L}_{\alpha\lambda}^{\lambda}%
+\mathcal{L}_{\rho\lambda}^{\lambda}\mathcal{L}_{\alpha\beta}^{\rho
}-\mathcal{L}_{\rho\beta}^{\lambda}\mathcal{L}_{\alpha\lambda}^{\rho}%
+\Lambda_{\lambda\rho}^{\lambda}\Lambda_{\beta\alpha}^{\rho}-\Lambda
_{\beta\rho}^{\lambda}\Lambda_{\lambda\alpha}^{\rho},\\
\overline{Q}_{\beta\alpha}  &  =\text{ }\partial_{\lambda}\text{ }%
\Lambda_{\alpha\beta}^{\lambda}-\partial_{\alpha}\text{ }\Lambda_{\lambda
\beta}^{\lambda}+\mathcal{L}_{\rho\lambda}^{\lambda}\Lambda_{\alpha\beta
}^{\rho}-\mathcal{L}_{\rho\alpha}^{\lambda}\Lambda_{\lambda\beta}^{\rho
}\text{{}}+\Lambda_{\lambda\rho}^{\lambda}\mathcal{L}_{\beta\alpha}^{\rho
}-\Lambda_{\alpha\rho}^{\lambda}\mathcal{L}_{\beta\lambda}^{\rho}.
\end{align}

Using the antisymmetric property of $\Lambda_{\alpha\beta}^{\lambda},$ two
scalar curvature are obtained from $P_{\alpha\beta}$ and $\overline{P}%
_{\beta\alpha}$
\begin{align}
P  &  =g^{\alpha\beta}P_{\alpha\beta}=R+\Lambda+2\nabla^{\alpha}S_{\alpha
},\label{V2}\\
\overline{P}  &  =g^{\alpha\beta}\overline{P}_{\beta\alpha}=R+\Lambda,
\end{align}
where $R$ is the Ricci scalar obtained by contracting the Ricci tensor%
\[
R_{\alpha\beta}=\partial_{\lambda}\mathcal{L}_{\alpha\beta}^{\lambda}%
-\partial_{\beta}\mathcal{L}_{\alpha\lambda}^{\lambda}+\mathcal{L}
_{\alpha\beta}^{\rho}\mathcal{L} _{\rho\lambda}^{\lambda}-\mathcal{L}
_{\rho\beta}^{\lambda}\mathcal{L} _{\alpha\lambda}^{\rho},
\]
and $S_{\alpha}=S_{\alpha\lambda}^{\lambda}$ is the torsion vector. The scalar
$\Lambda$ is defined by \cite{Clerc2}%
\begin{equation}
\Lambda=g^{\alpha\beta}\Lambda_{\alpha\lambda}^{\sigma}\Lambda_{\beta\sigma
}^{\lambda}. \label{V3}%
\end{equation}

In the case $S_{\alpha}=0,$ the scalar $\Lambda=P-R$ appears as a correction
of the curvature, and represents the contribution of a new tensor field
$\Lambda_{\beta\lambda}^{\varrho}$ which results from the immersion of the
manifold $V_{4}$ in $V_{8}$ $\cite{Clerc2}.$

One can also obtain two other scalar curvatures from $Q_{\alpha\beta}$ and
$\overline{Q}_{\beta\alpha}$
\begin{align}
Q &  =g^{\alpha\beta}Q_{_{\alpha\beta}}=g^{\alpha\beta}\left(  \Lambda
_{\beta\lambda}^{\varrho}S_{\varrho\alpha}^{\lambda}+\Lambda_{\alpha\lambda
}^{\varrho}S_{\varrho\beta}^{\lambda}\right)  ,\\
\overline{Q} &  =g^{\alpha\beta}\overline{Q}_{\beta\alpha}=Q.
\end{align}
The general field equations can be obtained from the general action
\cite{Clerc2}
\begin{equation}
S=\int\sqrt{-g}\left(  \frac{P+\overline{P}}{2}+\theta\frac{Q+\overline{Q}}%
{2}\right)  d^{4}x,\label{action}%
\end{equation}
where $g=\det g_{\mu\nu}$ and $\theta$ is a function of coordinates of
$V_{4}.$

These results have been obtained for the general case of a nonsymmetric
connection by Crumeyrolle and Clerc. In our work, we will consider the simple
case of vanishing torsion in $V_{4}$, and write the field equations and
discuss the consequences.

\section{Modified Einstein's equations}

Following the same steps as Crumeyrolle, where the four dimensional space-time
of general relativity is considered as a diagonal manifold $V_{4}$ with
vanishing torsion, all $S_{\rho\alpha}^{\lambda}=0$, immersed in an eight
dimensional hypercomplex manifold $V_{8}$. This implies that the connection
$\left(  \ref{D14}\right)  $ is reduced to the symmetric Levi-Civita connection%

\begin{equation}
\mathcal{L}_{\beta\gamma}^{\alpha}=\left\{  _{\beta\gamma}^{\alpha}\right\}
=\frac{1}{2}g^{\alpha\rho}\left(  \partial_{\beta}g_{\rho\gamma}%
+\partial_{\gamma}g_{\beta\rho}-\partial_{\rho}g_{\gamma\beta}\right)  ,
\label{V4}%
\end{equation}
and then the scalar curvatures defined in the last section are reduced to
\begin{equation}
P=\overline{P}=R+\Lambda,\text{ \ \ \ \ \ \ \ \ \ }Q=\overline{Q}=0,
\label{V5}%
\end{equation}
where $R$ is the scalar curvature of the symmetric connection $\left(
\ref{V4}\right)  $.

From Eqs. $\left(  \ref{D15}\right)  $ and $\left(  \ref{V3}\right)  $, the
scalar $\Lambda$ can be written in the form
\begin{equation}
\Lambda=6g^{\mu\nu}U_{\mu}U_{\nu}=6U^{2}. \label{Lambda}%
\end{equation}

In this case the new Einstein-Hilbert action is obtained from the action
$(\ref{action})$
\begin{equation}
S=\int\sqrt{-g}Pd^{4}x=\int\sqrt{-g}(R+\Lambda)d^{4}x. \label{V6}%
\end{equation}

In the absence of ordinary matter, the field equations are obtained using a
variational principle (it is important to remember that $\Lambda$ depends on
$g^{\mu\nu}$)%
\begin{equation}
\delta S=\int\sqrt{-g}d^{\text{ }4}x\text{ }\left[  R_{\mu\nu}-\frac{1}%
{2}g_{\mu\nu}R-\frac{1}{2}\Lambda g_{\mu\nu}+\frac{\partial\Lambda}{\partial
g^{\mu\nu}}\right]  \delta g^{\mu\nu}=0.
\end{equation}
Let us put $K_{\mu\nu}=\frac{\partial\Lambda}{\partial g^{\mu\nu}}$, then we
obtain the modified Einstein's equations
\begin{equation}
R_{\mu\nu}-\frac{1}{2}g_{\mu\nu}R-\frac{1}{2}\Lambda g_{\mu\nu}+K_{\mu\nu}=0.
\label{V21}%
\end{equation}
These equations are the field equations in vacuum because we haven't
introduced any matter term in the action. In this case, the two last terms in
Eq.$\left(  \ref{V21}\right)  $ can be considered as a source for dark energy
and the scalar $\Lambda$ is identified to be the cosmological constant, and
its expression is given by $\left(  \ref{Lambda}\right)  $.

Then, we deduce that the cosmological constant $\Lambda$ is space-time
dependent, a Lorentz invariant scalar and its sign depends on the nature of
the arbitrary four-vector $U$; i.e., For a time like vector $\Lambda>0$, and
for a space-like vector $\Lambda<0$ and for a light-like vector $\Lambda=0$.

It is important to note that the cosmological constant $\Lambda$ has a
geometrical origin due to the immersion of the space-time submanifold $V_{4}$
in an eight dimensional manifold. It is defined as a trace of the tensor
$\Lambda_{\alpha\lambda}^{\sigma}\Lambda_{\beta\sigma}^{\lambda}$, and the
tensor $\Lambda_{\alpha\lambda}^{\sigma}$ verifies the two equations
$(\ref{D12})~$and $(\ref{D122})$. In conclusion, the geometrical nature of the
vector $U$ allows to determine the sign of the cosmological constant.

From equations $(\ref{V21}),$ a new energy-momentum tensor for dark energy can
be defined by
\begin{equation}
T_{\mu\nu}^{\text{ }DE}=\frac{1}{8\pi G}\left(  -\frac{1}{2}\Lambda g_{\mu\nu
}-K_{\mu\nu}\right)  , \label{V22}%
\end{equation}
this tensor contains two terms while in the Einstein cosmological model it
contains only the term $(-\frac{\Lambda}{8\pi G}g_{\mu\nu})$. In order to
determine the terms $\Lambda$ and $K_{\mu\nu}$ we have to know the expression
of the arbitrary $4$-vector $U.$

The general Bianchi identities in $V_{8}$ are written as
\cite{Crum1,Crum2,Crum3,Clerc1,Clerc2}
\begin{equation}
\left(  \partial_{\nu}R_{jkl}^{i}-R_{nkl}^{i}\Gamma_{jm}^{n}+R_{jkl}^{n}%
\Gamma_{nm}^{i}\right)  dx^{k}\wedge dx^{l}\wedge dx^{m}=0.
\end{equation}
By restriction in the diagonal submanifold $V_{4}$ ($x^{\mu^{\ast}}=0$)$,$ and
contraction one can obtain \cite{Clerc2}%
\begin{equation}
\nabla^{\nu}\left(  R_{\mu\nu}-\frac{1}{2}g_{\mu\nu}R\right)  =\nabla^{\nu
}\left(  2U_{\nu}U_{\mu}+g_{\mu\nu}U^{\text{ }^{2}}\right)  +U^{\sigma}\left(
\nabla_{\sigma}U_{\mu}-\nabla_{\mu}U_{\sigma}\right)  .\label{bian}%
\end{equation}
The appearance of the terms in right hand side of the last equation is due to
the immersion, these terms have to vanish in the case of a Levi-Civita
connection, then
\begin{equation}
\nabla^{\nu}\left(  2U_{\nu}U_{\mu}+g_{\mu\nu}U^{2}\right)  +U^{\sigma}\left(
\nabla_{\sigma}U_{\mu}-\nabla_{\mu}U_{\sigma}\right)  =0,
\end{equation}
which can be simplified to
\begin{equation}
\nabla_{\alpha}\left(  U^{2}U^{\alpha}\right)  =0.\label{consv}%
\end{equation}
In section six, we determine the four vector $U$ as a solution to this
equation and deduce the expression of the cosmological constant.

\section{Spherical solution}

The static spherically symmetric space-time metric is given by
\begin{equation}
ds^{2}=-e^{2\mu(r)}dt^{2}+e^{2\nu(r)}dr^{2}+r^{2}(d\theta^{2}+\sin^{2}\theta
d\phi^{2}), \label{s1}%
\end{equation}
where $\mu,\nu$ are functions of $r.$

Using the field equations in vacuum $\left(  \ref{V21}\right)  ,$ the general
solution for the metric $\left(  \ref{s1}\right)  $ is given by%

\begin{equation}
ds^{2}=-\exp\left\{  \int\left[  \frac{\frac{\Lambda}{2}r^{2}-r^{2}K_{_{rr}%
}-a_{1}-a_{2}}{r\left(  1+a_{1}+a_{2}\right)  }\right]  dr\right\}
dt^{2}+\frac{1}{1+a_{1}+a_{2}}dr^{2}+r^{2}(d\theta^{2}+\sin^{2}\theta
d\phi^{2}), \label{met}%
\end{equation}
where%
\begin{equation}
a_{1}=\frac{1}{2r}\int\Lambda r^{2}dr,\ \text{\ \ \ \ }a_{2}=\frac{1}{r}\int
r^{2}K_{_{00}}dr.
\end{equation}

In the particular case where $\Lambda$ is constant, i.e. $\delta\Lambda=0$ and
$K_{_{\mu\nu}}=0$, the above coefficients are reduced to $a_{1}=\frac{\Lambda
}{6}r^{2}$ and $a_{2}=0$ and the metric (\ref{met}) is reduced to the de
Sitter or anti-de Sitter metric
\begin{equation}
ds^{2}=-\left(  1+\frac{\Lambda}{6}r^{2}\right)  dt^{2}+\frac{1}%
{1+\frac{\Lambda}{6}r^{2}}dr^{2}+r^{2}(d\theta^{2}+\sin^{2}\theta d\phi^{2}),
\label{sitter}%
\end{equation}
we note that for a time like vector $U$, i.e. $\Lambda>0$, the metric
(\ref{sitter}) corresponds to an anti-de Sitter space. For a space like vector
$U$, i.e. $\Lambda<0$, the metric (\ref{sitter}) corresponds to a de Sitter
space. While for a light-like vector $U$, i.e. $\Lambda=0$, corresponds to a
flat space.

\section{Modified Friedmann equations}

Let us introduce the flat Friedmann-Robertson-Walker metric
\begin{equation}
ds^{2}=-dt^{2}+a^{2}(t)\left[  dr^{2}+r^{2}d\theta^{2}+r^{2}\sin^{2}\theta
d\phi^{2}\right]  ,
\end{equation}
where $a(t)$ is the scale factor. Using the field equations $($\ref{V21}$)$ ,
we obtain the Friedmann equations\qquad{}\qquad{}\qquad{}\
\begin{equation}
\left(  \frac{\overset{\cdot}{a}}{a}\right)  ^{2}=\frac{\Lambda}{6}%
-\frac{K_{0}^{0}}{3}, \label{F3}%
\end{equation}%
\begin{align}
\left(  \frac{\overset{\cdot\cdot}{a}}{a}\right)   &  =\frac{\Lambda}{6}%
+\frac{K_{0}^{0}}{6}+\frac{K_{r}^{r}}{2},\nonumber\\
\left(  \frac{\overset{\cdot\cdot}{a}}{a}\right)   &  =\frac{\Lambda}{6}%
+\frac{K_{0}^{0}}{6}+\frac{K_{\theta}^{\theta}}{2},\nonumber\\
\left(  \frac{\overset{\cdot\cdot}{a}}{a}\right)   &  =\frac{\Lambda}{6}%
+\frac{K_{0}^{0}}{6}+\frac{K_{\phi}^{\phi}}{2}.
\end{align}
The last three equations give
\begin{equation}
K_{r}^{r}=K_{\theta}^{\theta}=K_{\phi}^{\phi},
\end{equation}
then, we obtain the spacial equation%
\begin{equation}
\left(  \frac{\overset{\cdot\cdot}{a}}{a}\right)  =\frac{\Lambda}{6}%
+\frac{K_{0}^{0}}{6}+\frac{K_{r}^{r}}{2}. \label{F8}%
\end{equation}
From Eq. $\left(  \ref{F3}\right)  $, we obtain the condition
\begin{equation}
\frac{\Lambda}{6}-\frac{K_{0}^{0}}{3}>0, \label{F10}%
\end{equation}
which gives a positive energy density for the dark energy
\begin{equation}
\rho=\frac{1}{8\pi G}\left(  \frac{1}{2}\Lambda-K_{0}^{0}\right)  >0.
\end{equation}
We also define the pressure of the dark energy by%
\begin{equation}
p_{_{DE}}=\frac{1}{8\pi G}\left(  -\frac{1}{2}\Lambda-K_{r}^{r}\right)  .
\end{equation}
Integrating equation $\left(  \ref{F3}\right)  $, we obtain
\begin{equation}
a\left(  t\right)  =a\left(  t_{0}\right)  \exp\left[  \int\sqrt{\frac
{\Lambda}{6}-\frac{K_{0}^{0}}{3}}\text{ }dt\right]  . \label{F12}%
\end{equation}
This is like the case of the de Sitter space: it describes an empty
exponentially expanding space. The differences from the de Sitter case are the
presence of the term $K_{0}^{0}$ and here $\Lambda$ is not constant.

Using Eqs.$\left(  \ref{F3}\right)  $ and $\left(  \ref{F8}\right)  $ we
obtain the expression of the deceleration parameter $q$%
\begin{equation}
q=-\frac{\overset{\cdot\cdot}{a}a}{\overset{\cdot}{a}^{2}}=\left[  \frac
{\frac{\Lambda}{6}+\frac{K^{0}_{0}}{6}+\frac{K^{r}_{r}}{2}}{ \frac{\Lambda}%
{6}-\frac{K^{0}_{0}}{3} }\right]  ,
\end{equation}
and for the equation of state $\omega=\frac{p_{_{_{DE}}}}{\rho_{_{_{DE}}}},$
we have
\begin{equation}
\omega=-\frac{\frac{1}{2}\Lambda+K^{r}_{r}}{\frac{1}{2}\Lambda-K^{0}_{0}},
\end{equation}
for the particular case $K^{0}_{0}=-K^{r}_{r}$, we obtain the well known
result $\omega=-1.$

We remark that the deceleration parameter $q$ vanishes for
\begin{equation}
\Lambda= -\left(  K^{0}_{0}+3K^{r}_{r}\right)  ,
\end{equation}
which gives%
\begin{equation}
\omega=-\frac{1}{3}.
\end{equation}
For an accelerated universe $\left(  q<0\right)  $, we must have the condition%
\begin{equation}
\Lambda>-\left(  K^{0}_{0}+3K^{r}_{r}\right)  , \label{cond02}%
\end{equation}
which gives with Eq.$\left(  \ref{F10}\right)  $, the well known condition
$\omega<-\frac{1}{3}.$

From the conditions $\left(  \ref{F10}\right)  $ and $\left(  \ref{cond02}%
\right)  $ we obtain%
\begin{equation}
-\frac{1}{2}\Lambda-K^{r}_{r}<0,
\end{equation}
which implies that the pressure of the dark energy is negative
\begin{equation}
p_{_{DE}}=\frac{1}{8\pi G}\left(  -\frac{1}{2}\Lambda-K^{r}_{r}\right)  <0.
\end{equation}
At the end, we note that for a decelerating phase $\left(  q>0\right)  $, we
have
\begin{equation}
\Lambda<-\left(  K^{0}_{0}+3K^{r}_{r}\right)  ,
\end{equation}
which gives with Eq.$\left(  \ref{F10}\right)  $, the condition $\omega
>-\frac{1}{3}$. As we saw, what we have given until now is a general frame
work, and one has to give an expression for the four vector $U,$ which can be
obtained from the condition $\left(  \ref{consv}\right)  .$

\section{Cosmological constant as a function of the scale factor}

As an application, we propose a solution to the equation $\left(
\ref{consv}\right)  $ in the form \cite{Crum2,Clerc2}
\begin{equation}
U_{\mu}\propto\sqrt{\left(  -g\right)  ^{-\alpha}}u_{\mu}, \label{form1}%
\end{equation}
where $\alpha$ is a constant, $u_{\mu}$ is a unit vector which verifies
$g^{\mu\nu}u_{\mu}u_{\nu}=1,$ and $g=\det g_{\mu\nu}.$

From this solution we obtain the cosmological constant
\[
\Lambda=6g^{\mu\nu}U_{\mu}U_{\nu}\propto6\left(  -g\right)  ^{-\alpha}.
\]
Using the flat Friedmann-Robertson-Walker (FRW) metric in cartesian
coordinates
\begin{equation}
ds^{2}=-dt^{2}+a^{2}(t)\left[  dx^{2}+dy^{2}+dz^{2}\right]  ,
\end{equation}
i.e, $g_{\mu\nu}=diag\left[  -1,a^{2},a^{2},a^{2}\right]  ,$ which gives for
the determinant
\begin{equation}
(-g)=a^{6}(t),
\end{equation}
and finally the cosmological constant becomes time dependent or scale factor
dependent
\begin{equation}
\Lambda\propto6a^{-6\alpha}(t). \label{form2}%
\end{equation}
This is a general expression of the time dependent cosmological constant.

This scale factor dependent cosmological constant may give theoretically an
explanation of why it has decayed from a large value in the early universe to
have small one today. Particle and cosmology theories suggest that as the
universe expanded and cooled different phases took place (symmetry breaking)
and the energy density associated to the cosmological constant decreased from
an initially large value which is supported by the inflationary scenario
\cite{Weinberg,Guth} and attained a small value today. Although we did not
give numerical values, the evolving cosmological constant (\ref{form2}) is
interesting, the energy associated to it is decreasing with an increasing
scale factor (the expanding). For physical reasons which motivate a decaying
cosmological constant, we refer the reader to Refs.
\cite{Taha1,Chen,Lopez,Taha,Joseph,Mbonye}, our approach here in deriving this
formula is different from those references, in fact we gave a pure geometric
origin of the cosmological constant.

Using the expressions $U_{\mu}=\sqrt{\left(  -g\right)  ^{-\alpha}}u_{\mu}$
and $\Lambda=6g^{\mu\nu}U_{\mu}U_{\nu}=$ $6(-g)^{-\alpha}$ we obtain the
tensor
\begin{align*}
K_{\mu\nu}  &  =\frac{\partial\Lambda}{\partial g^{\mu\nu}}=6U_{\mu}U_{\nu
}+6g^{\alpha\beta}\frac{\partial U_{\alpha}}{\partial g^{\mu\nu}}U_{\beta
}+6g^{\alpha\beta}U_{\alpha}\frac{\partial U_{\beta}}{\partial g^{\mu\nu}}\\
&  =\Lambda\left(  u_{\mu}u_{\nu}+\alpha g_{\mu\nu}\right)  ,
\end{align*}
and the energy momentum tensor (\ref{V22}) becomes
\begin{equation}
T_{\mu\nu}^{\text{ }DE}=-6a^{-6\alpha}(t)\left[  \left(  \alpha+\frac{1}%
{2}\right)  g_{\mu\nu}+u_{\mu}u_{\nu}\right]  , \label{de}%
\end{equation}
with $8\pi G=1.$ This gives the density and the pressure%
\begin{equation}
\rho^{DE}=-T_{0}^{0DE}=3\left(  2\alpha-1\right)  a^{-6\alpha},\text{
\ \ \ \ \ \ }p^{DE}=T_{i}^{iDE}=-3\left(  2\alpha+1\right)  a^{-6\alpha}.
\label{ro}%
\end{equation}
We note here that the energy density is positive only for $\alpha>\frac{1}%
{2},$ and this is a condition for the value of $\alpha.$

In this model, the two Friedmann equations in the case of dark energy
dominated universe become%
\begin{equation}
\left(  \frac{\overset{\cdot}{a}}{a}\right)  ^{2}=a^{-6\alpha}\left(
2\alpha-1\right)  \text{ and \ }\left(  \frac{\overset{\cdot\cdot}{a}}%
{a}\right)  =2a^{-6\alpha}\left(  \alpha+1\right)  ,\text{ } \label{frr1}%
\end{equation}
as we see, we have an accelerating phase, since $\alpha+1>0$ when
$\alpha>\frac{1}{2}.$ This also can be seen from the deccelerating parameter%
\[
q=-\frac{\overset{\cdot\cdot}{a}a}{\overset{\cdot}{a}^{2}}=-\frac{2\left(
\alpha+1\right)  }{\left(  2\alpha-1\right)  }<0\text{ \ \ for all }%
\alpha>\frac{1}{2}.
\]
We can integrate the first Friedmann equation in (\ref{frr1}) and obtain
\begin{equation}
a\left(  t\right)  =\left(  3\alpha\right)  ^{\frac{1}{3\alpha}}\left(
2\alpha-1\right)  ^{\frac{1}{6\alpha}}t^{\frac{1}{3\alpha}}, \label{scale}%
\end{equation}
where we have put $a\left(  t=0\right)  =0$. When substituted in the energy
density (\ref{ro}) it gives%
\begin{equation}
\rho^{DE}=\frac{1}{3\alpha^{2}}t^{-2}, \label{rot0}%
\end{equation}
i.e, $\rho^{DE}$ decays with time as%
\begin{equation}
\rho^{DE}\propto t^{-2}. \label{rot}%
\end{equation}
This time behavior of the dark energy density \cite{choi} is the same as the
evolution of ordinary matter in the case of matter dominated era $\rho\propto
a^{-3}=t^{-2}$ (because in that era, $a\left(  t\right)  =t^{\frac{2}{3}}$ ).
We think this is interesting, it may give an explanation of the coincidence
problem, i.e " why the energy density in matter is close to the vacuum energy
today? ".

From equation (\ref{scale}), we obtain the Hubble parameter
\[
H=\frac{\overset{\cdot}{a}}{a}=\frac{1}{3\alpha}t^{-1},
\]
this gives the age of the universe (when $H$ is given)%
\begin{equation}
t=\frac{1}{3\alpha H}\sim\frac{2}{3H}\text{ \ \ for }\alpha>\frac{1}{2},
\label{age}%
\end{equation}
the value $\frac{2}{3H}$ is the age of the universe obtained in the case of
ordinary matter era and is numerically $\frac{2}{3h}.10^{10}$ years, with
$0,5\leq h\leq1.$

Now let us study the behavior of the ordinary matter and radiation. In this
case the energy momentum tensor of the matter (radiation) will appear in the
Einstein's field equations as a perfect fluid $T_{\mu\nu}^{\text{ }m,r}%
=(\rho+p)u_{\mu}u_{\nu}+pg_{\mu\nu},$ and due to dark energy the Friedmann
equations become \qquad{}\qquad{}\qquad{}\
\begin{equation}
\left(  \frac{\overset{\cdot}{a}}{a}\right)  ^{2}=\frac{8\pi G}{3}%
\rho+a^{-6\alpha}\left(  2\alpha-1\right)  \text{ \ \ \ \ and \ \ \ }\left(
\frac{\overset{\cdot\cdot}{a}}{a}\right)  =-\frac{4\pi G}{3}\left(
\rho+3p\right)  +2a^{-6\alpha}\left(  \alpha+1\right)  , \label{Fried}%
\end{equation}
here $\rho$ and $p$ may be the density and pressure of matter or radiation or
both of them, and the appearance of the second terms in both equations is due
to $\rho^{DE}$ and $p^{DE}$.

The conservation law $\nabla_{\mu}T_{\nu}^{\text{ }\mu}=0$, where $T_{\mu\nu
}=T_{\mu\nu}^{\text{ }m,r}+T_{\mu\nu}^{\text{ }DE}$, becomes%
\begin{equation}
\overset{\cdot}{\rho}+3\frac{\overset{\cdot}{a}}{a}\left(  \rho+p\right)
=18\left(  3-2\alpha\right)  \frac{\overset{\cdot}{a}}{a^{6\alpha+1}}.
\label{cons}%
\end{equation}
This is the new conservation law with the new term in the right hand side as a
consequence of the time dependence of the density and pressure of dark energy
described in this model. For matter $\left(  p=0\right)  $ or for radiation
$\left(  p=\frac{1}{3}\rho\right)  $, equation (\ref{cons}) becomes
respectively
\begin{equation}
\overset{\cdot}{\rho}^{m}+3\frac{\overset{\cdot}{a}}{a}\rho^{m}=18\left(
3-2\alpha\right)  \frac{\overset{\cdot}{a}}{a^{6\alpha+1}}\text{ \ \ \ and
\ \ }\overset{\cdot}{\rho}^{r}+4\frac{\overset{\cdot}{a}}{a}\rho^{r}=18\left(
3-2\alpha\right)  \frac{\overset{\cdot}{a}}{a^{6\alpha+1}}. \label{rm}%
\end{equation}
The solutions of these equations give a correction to the behavior of the
energy density of matter and radiation%
\begin{equation}
\rho^{m}\sim A_{1}a^{-3}(t)+A_{2}a^{-6\alpha}(t)\text{ \ and \ }\rho^{r}\sim
B_{1}a^{-4}(t)+B_{2}a^{-6\alpha}(t), \label{sol}%
\end{equation}
where $A_{1}$, $A_{2}$, $B_{1}$ and $B_{2}$ are constants.

We note here that the solutions (\ref{sol}) are different from the standard
ones, $\rho^{m}\sim a^{-3}$ and $\rho^{r}\sim a^{-4}$, by the presence of the
additional term $a^{-6\alpha}(t)$ due the time dependence of the dark energy.
These solutions combined with Friedmann equations (\ref{Fried}) will give the
time evolution of the scale factor $a(t)$.

\section{Conclusion}

In this paper, we have used Crumeyrolle's results on hypercomplex manifold
where the four dimensional space-time of general relativity is considered as a
submanifold immersed in an eight dimensional hypercomplex manifold. In the
case of symmetric connection, we have seen that the cosmological constant
$\Lambda$ appears naturally in Einstein's equations and its expression is
given as a norm of a four-vector $U$. Then, the cosmological constant can be
positive, negative or null. A new energy momentum tensor of the dark energy is
obtained which depends on the cosmological constant and its first derivatives
with respect to the metric. In the first application, the spherical solution
of the Einstein's equations is obtained. In the second, we have obtained the
modified Friedmann equation in the standard flat Friedmann-Robertson-Walker
metric, and found that the equation of state depends on $\Lambda$ and its
first derivative with respect to the metric $\omega=\omega(\Lambda
,\frac{\partial\Lambda}{\partial g^{\mu\nu}}).$ At the end, a condition on
$\Lambda$ was deduced for an accelerating universe which is equivalent to the
well known condition $\omega<-\frac{1}{3}$. For a particular case of the four
vector $U$, we deduced a decaying cosmological constant $\Lambda\propto
a(t)^{-6\alpha}$, which in turn modifies the behaviors of ordinary matter and radiation.

\section*{Acknowledgements}

A.B thanks Prof. Peter O. Hess (UNAM-Mexico) and Prof. Walter Greiner
(FIAS-Frankfurt) for drawing my attention to the works of Crumeyrolle, and
Profs. J.P. Provost (Univ. Nice) and B. Chauvineau (Observatoire C\^ote d'Azur
Nice) for the usefull discussion particularly to obtain Eq. (\ref{consv}).
This work was supported by the Algerian Ministry of Higher Eduction and
Scientific Research (MESRS).

\end{document}